\documentclass[water,commentary,submit,moreauthors,10pt,a4paper]{arxiv} 
\usepackage{mathpazo}
\usepackage[T1]{fontenc}
\usepackage{amsmath}
\usepackage{url}
\usepackage{hyperref}
\usepackage{setspace}
\usepackage{siunitx}
\firstpage{1} 
\Title{Commentary on `Modified MMF (Morgan--Morgan--Finney) model for evaluating effects of crops and vegetation cover on soil erosion' by Morgan and Duzant (2008)}

\Author{Kwanghun Choi $^{1}$*, Bernd Huwe $^{2}$, and Bj\"orn Reineking $^{1,3,4}$}

\address{%
    $^{1}$ \quad Department of Biogeographical Modelling, Bayreuth Center of Ecology and Environmental Research (BayCEER), University of Bayreuth, Universit\"atstra{\ss}e 30, 95440, Bayreuth, Germany\\
$^{2}$ \quad Department of Soil Physics, Bayreuth Center of Ecology and Environmental Research (BayCEER), University of Bayreuth, Universit\"atstra{\ss}e 30, 95440, Bayreuth, Germany\\
$^{3}$ \quad Irstea, UR EMGR, 2 rue de la Papeterie-BP 76, 38402, St-Martin-d'H{\`e}res, France\\
$^{4}$ \quad Department of Particle Physics, Astrophysics, Geosciences, Ecology, and Environment, Universit{\'e} Grenoble Alpes, 38402, Grenoble, France}

\corres{Correspondence: kwanghun.choi@yahoo.com; bt302546@uni-bayreuth.de}

\abstract{The Morgan--Morgan--Finney (MMF) model is a widely used semi-physically based soil erosion model that has been tested and validated in various land use types and climatic regions. The latest version of the model, the modified MMF (MMMF) model, improved its conceptual physical representations through several modifications of the original model. However, the MMMF model has three problematic parts to be corrected: 1) the effective rainfall equation, 2) the interflow equation, and 3) the improperly normalized C-factor of the transport capacity equation. In this commentary, we identify and correct the problematic parts of the MMMF model, which should result in more accurate estimations of runoff and soil erosion rates.
}
\keyword{effective rainfall, interflow, C-factor, MMF, soil erosion model, error correction}

\begin{document}
\section{Introduction}
The Morgan--Morgan--Finney (MMF) model \citep{Morgan1984} is a semi-physically based model used to estimate the amount of annual runoff and soil eroded from a field or a catchment. Similar to physically based models such as SWAT \citep{SWAT2009}, EUROSEM \citep{EUROSEM1998}, LISEM \citep{LISEM1996}, and WEPP \citep{WEPP1989}, the MMF model has the properties of both physically based and empirical models and provides an in-depth understanding of soil erosion processes by using physical concepts.
Moreover, the MMF model, similarly to empirical models such as USLE \citep{USLE1978} and RUSLE \citep{Renard1991}, maintains a conceptual simplicity by using semi-empirical relationships and does not require the excessive parameters and computing resources \citep{Morgan1984,Morgan2001,Morgan2008,Lilhare2014}.
%Moreover, the MMF model maintains a conceptual simplicity by using semi-empirical relationships and does not require the excessive parameters and computing resources needed for empirical models such as USLE \citep{USLE1978} and RUSLE \citep{Renard1991} \citep{Morgan1984,Morgan2001,Morgan2008,Lilhare2014}.
For this reason, the MMF and the revised MMF (RMMF) \citep{Morgan2001} models have been applied and validated in a variety of climatic regions and land use types \citep{Morgan1984, DeJong1999, Morgan2001, Vigiak2005, LopezVicente2008, Pandey2009, Li2010MMF, Feng2014, Tesfahunegn2014, Vieira2014}.
In the latest version of the MMF, the modified MMF (MMMF) \citep{Morgan2008}, hydrological processes were improved by considering the slope angle in the calculation of effective rainfall and introducing interflow processes.
In addition, the soil erosion processes of the MMMF model were improved by introducing gravitational deposition process, generalizing the effect of ground surface on sediment deposition and transportation, and considering the characteristics of each soil particle type \citep{Morgan2008, Lilhare2014}. 
These modifications allow the MMMF model to consider physical aspects of terrain and soil surface conditions more effectively than previous versions of the MMF model.
However, we argue that errors persist in effective rainfall, interflow, and transport capacity equations, which ultimately affects the model outputs in certain conditions. 
Despite these errors, however, the MMMF model was implemented and used in several studies without apparent consideration of the problematical parts (i.e., \citet{Setiawan2012} and \citet{Lilhare2014}).
In addition, one of the problematical parts of the MMMF model, the problematic slope adjustment factor of effective runoff, was used in the Modified-RMMF-2014 model of \citet{LopezVicente2015}.
The objective of the present study is to identify and correct the problematic terms concerning:
\begin{enumerate}[leftmargin=*,labelsep=3mm]
    \item a trigonometric error in the calculation of effective rainfall ($R\!f$),
    \item a quantity estimation error in the calculation of interflow ($IF$), and
    \item an improperly normalized C-factor in the transport capacity equation ($TC$).
\end{enumerate}
\section{Problematic parts of the MMMF model}
\subsection{Trigonometric error in the calculation of effective rainfall}
\begin{figure}[H]
    \includegraphics{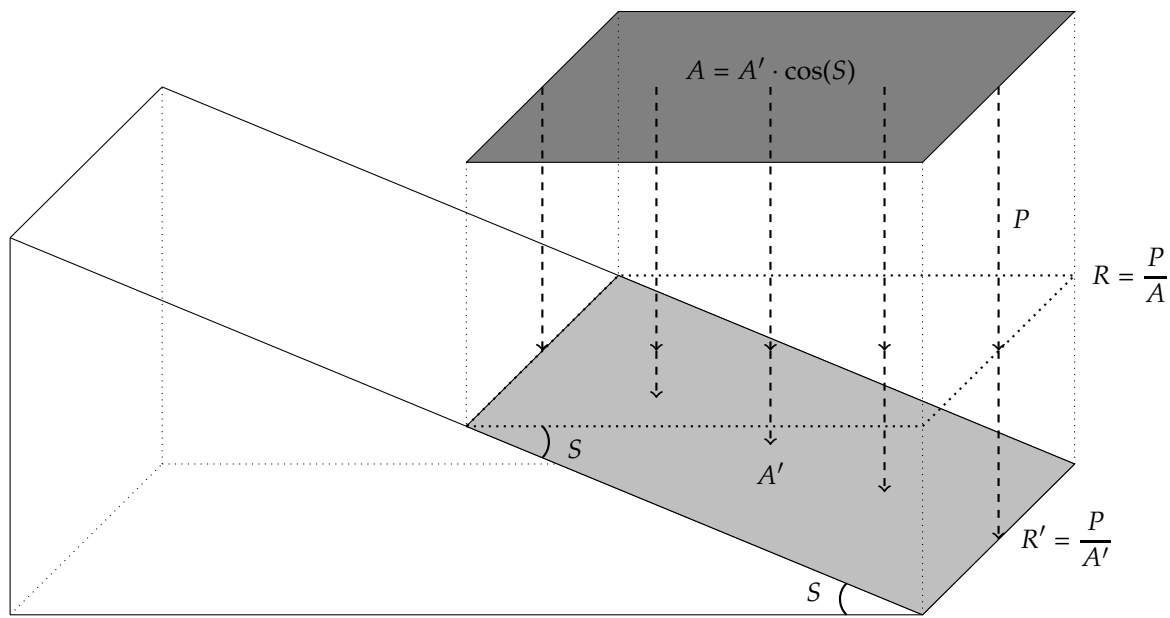}
    \caption{Conceptual representation of a hillslope with a slope angle of $S$ (\si{\degree}). $A$ (\si{\square\m}) is the area of a horizontal plane, and $A'$ (\si{\square\m}) is the projected area of $A$ on a hillslope.%
    Because the volume of rainfall ($P$; \si{\liter}) is the same for both $A$ and $A'$, the rainfall per unit area for both areas are $\frac{P}{A}$ (= $R$; \si{\mm}) and $\frac{P}{A'}$ (= $R'$; \si{\mm}). % 
    From the trigonometric rule, $R'$ is equal to $R \cdot \cos{(S)}$.}%
    \label{fig:Rf}
\end{figure}

The MMMF model represents the catchment through several interconnected elements, each of which has a uniform slope, land cover, and soil type. 
In the MMF model, effective rainfall is the primary source of the hydrological processes, which regulate surface runoff and the soil erosion processes.
The MMMF model calculates effective rainfall ($R\!f_{\mathrm{MMMF}}$; \si{\mm}) while considering the slope of a given element by using the following equation (\citet{Morgan2008}, eq. (1)):
\begin{equation}
    R\!f_{\mathrm{MMMF}} = R \cdot (1-PI) \cdot \frac{1}{\cos(S)},
\end{equation}
where $R$ (\si{\mm}) is the mean annual rainfall, $PI$ is the area proportion of the permanent interception of rainfall, and $S$ (\si{\degree}) is the slope of an element.
\newline
However, we argue that in order to calculate effective rainfall correctly, $\cos(S)$ should be used as a sloping adjustment factor rather than $\frac{1}{\cos(S)}$ as described in \citet{Sharon1980} and \citet{Tani1997}. 
We demonstrate our claims through mathematical proof and in Figure~\ref{fig:Rf}.
Let us consider an element on a hill slope with an angle of $S$ (\si{\degree}).
Assuming that the area in the horizontal plane is $A$ (\si{\square\m}) and its projected area on the element is $A'$ (\si{\square\m}), the trigonometric relationship between $A$ and $A'$ is
\begin{equation}
    A = A' \cdot \cos(S) \label{eq:A}.
\end{equation}
Because the total volume of rainfall ($P$; \si{\liter}) is the same for both $A$ and $A'$ (Figure~\ref{fig:Rf}), the amount of rainfall per unit area for $A$ and $A'$ can be calculated as
\begin{align}
    &R\phantom{'}=\frac{P}{A} \label{eq:R}\\ 
    &R'=\frac{P}{A'} \label{eq:Rprime}.
\end{align}
From the equations ~(\ref{eq:A}), ~(\ref{eq:R}), and ~(\ref{eq:Rprime}), we can estimate the rainfall per unit surface area on a hillslope ($R'$; \si{\mm}) with the rainfall of the area ($R$; \si{\mm}).
\begin{equation}
R' = \frac{P}{A'} = \frac{P}{A} \cdot \cos(S) = R \cdot \cos(S)
\end{equation}
If the element has areas with permanent interception of rainfall ($PI$), the effective rainfall per unit surface area ($R\!f_{\mathrm{corrected}}$) should be calculated as
\begin{equation}
    R\!f_{\mathrm{corrected}} = R' \cdot (1-PI) = R \cdot (1-PI) \cdot \cos(S).
\end{equation} 
Thus, the slope adjustment factor should be $\cos(S)$ rather than $\frac{1}{\cos(S)}$ in order to calculate effective rainfall ($R\!f$) considering the slope.
\subsubsection{Consequence of the error in calculating effective rainfall}
\begin{figure}[H]
    \centering
    \includegraphics{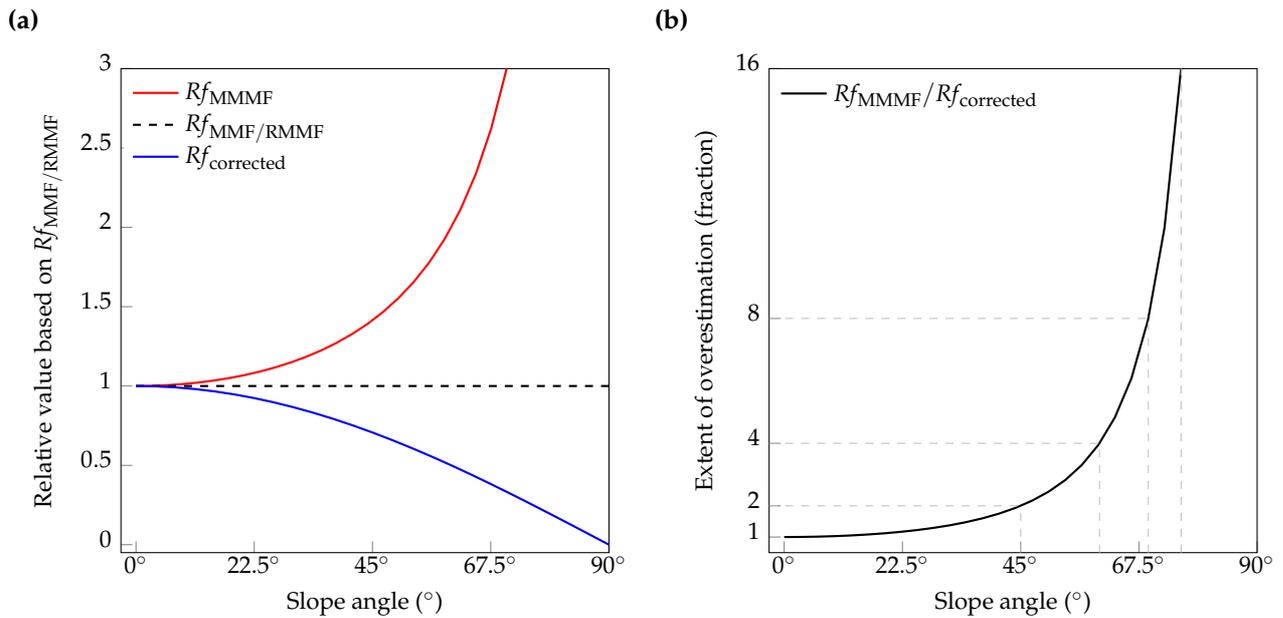}
    \caption{\textbf{(a)} Relative values of effective rainfall based on the slope invariant effective rainfall of the Morgan--Morgan--Finney (MMF) and revised MMF (RMMF) models ($R\!f_{\mathrm{MMF/RMMF}}$) as a function of slope angle.
    \textbf{(b)} Extent of overestimation of effective rainfall from the MMMF model ($R\!f_{\mathrm{MMMF}}$) to the corrected value ($R\!f_{\mathrm{corrected}}$) as a function of slope angle.
    The dashed gray lines in \textbf{(b)} indicate the extent of overestimation and the corresponding slope angles.
    } 
    \label{fig:ReffComp}
\end{figure}

Owing to the problematic slope adjustment factor of the effective rainfall suggested by Morgan and Duzant (2008), the model overestimates the effective rainfall when the slope of an element increases, as shown in Figure~\ref{fig:ReffComp}.
Considering that the MMF model and its variants have been applied in mountainous areas with steep hillslopes, as listed in Table~\ref{tab:Reff}, the MMMF model has a high risk of overestimation of effective rainfall.
\begin{table}[H]
\centering
\caption{Previous studies in which the Morgan--Morgan--Finney (MMF) model and its variants, the revised MMF (RMMF) and the modified MMF (MMMF), were applied to steep hillslopes}
\label{tab:Reff}
\begin{tabular}{lllll}
\toprule
Source                  & Year & Model& Site                            & Slope condition\\
\midrule
\citet{Feng2014}        & 2014 & RMMF & Guzhou catchment, China         & \textgreater \SI{55}{\degree} (\SI{15.8}{\percent} of total area) \\
                        &      &      &                                 & \SI{35}{\degree} - \SI{55}{\degree} (\SI{45.3}{\percent} of total area)\\
                        &      &      &                                 & \SI{25}{\degree} - \SI{35}{\degree} (\SI{20.7}{\percent} of total area)\\
                        &      &      &                                 & \SI{15}{\degree} - \SI{25}{\degree} (\SI{9.2}{\percent} of total area)\\
                        &      &      &                                 & \SI{0}{\degree} - \SI{15}{\degree} (\SI{9.0}{\percent} of total area)\\
\citet{Lilhare2014}     & 2014 & MMMF & Gamber watershed, India         & Steep topographic gradient \\
\citet{Tesfahunegn2014} & 2014 & RMMF & Mai-Negus catchment, Ethiopia   & Maximum slope of \SI{73}{\degree}\\
\citet{Setiawan2012}    & 2012 & MMMF & Kejajar Sub-district, Indonesia & \textgreater \SI{31.0}{\degree} (\SI{42.9}{\percent} of total area)\\
                        &      &      &                                 & \SI{16.7}{\degree} - \SI{31.0}{\degree} (\SI{28.3}{\percent} of total area)\\
                        &      &      &                                 & \SI{8.5}{\degree} - \SI{16.7}{\degree} (\SI{16.1}{\percent} of total area)\\
                        &      &      &                                 & \SI{0.0}{\degree} - \SI{8.5}{\degree} (\SI{11.7}{\percent} of total area)\\
\citet{Li2010MMF}       & 2010 & RMMF & Zuli River Basin, China         & Maximum slope of \SI{45}{\degree}\\
\citet{Pandey2009}      & 2009 & RMMF & Dikrong river basin, India      & \textgreater \SI{45}{\degree} (\SI{21.0}{\percent} of total area)\\
                        &      &      &                                 & \SI{36}{\degree} - \SI{45}{\degree} (\SI{1.8}{\percent} of total area)\\
                        &      &      &                                 & \SI{16}{\degree} - \SI{35}{\degree} (\SI{12.8}{\percent} of total area)\\
                        &      &      &                                 & \textless \SI{15}{\degree} (\SI{64.4}{\percent} of total area)\\
\citet{Vigiak2005}      & 2005 & RMMF & Kwalei catchment, Tanzania      & \textgreater \SI{11.3}{\degree} (\SI{50}{\percent} of total area)\\
                        &      &      & Gikuuri catchment, Kenya        & \SI{1.1}{\degree} - \SI{28.8}{\degree} (mean: \SI{10.2}{\degree})\\
\bottomrule
\end{tabular}
\end{table}
In the case of \citet{Setiawan2012}, the MMMF model overestimated effective rainfall by at least \SI{136}{\percent} of the corrected value at areas with a slope greater than \SI{31.0}{\degree}, which account for most of the research site.
If the MMMF model had been applied to the site of \citet{Tesfahunegn2014}, it would have overestimated effective rainfall 3.4 times more than that in the previous versions which do not consider the slope and 11.7 times more than the corrected value for the area with the maximum slope angle.
According to the sensitivity analysis of \citet{Morgan2008}, the model outputs of surface runoff and soil loss are highly sensitive to effective rainfall.
Moreover, the overestimation can be greater on downslope elements where the overestimated runoff from upslope accumulates.
Therefore, the trigonometric error in the calculation of effective rainfall may lead to incorrect results to a significant degree if the model is applied to a large watershed with steep slopes.
\subsection{Quantity estimation error in calculating interflow}
\subsubsection{Incorrect formula in the interflow equation}
\begin{figure}[H]
    \includegraphics{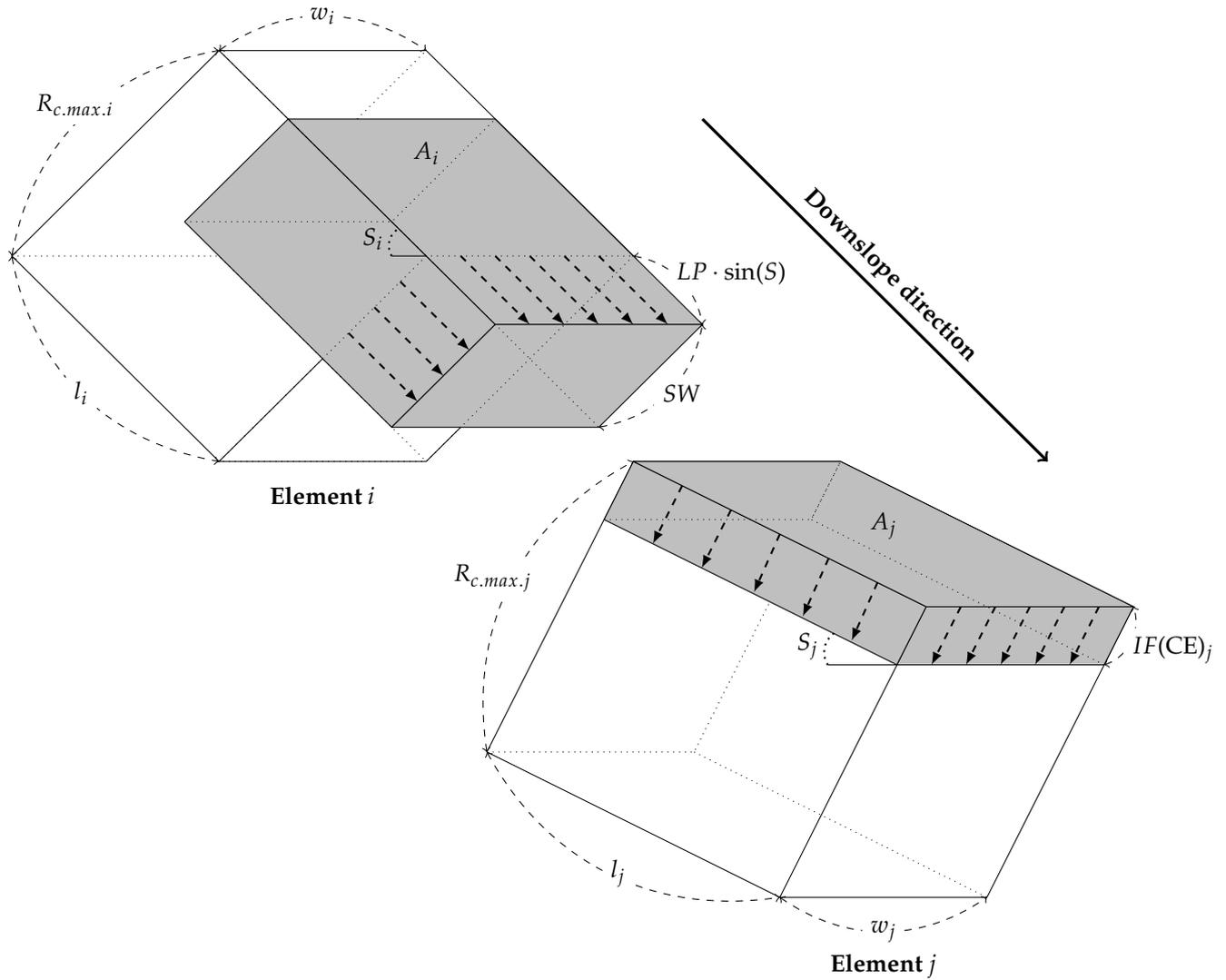}
    \caption{Conceptual representation of interflow between two adjacent elements $i$ and $j$.
        Here, $A$, $w$, $l$, $R_{c.max}$, and $S$ represent the surface area, width, length, maximum soil moisture storage capacity, and slope angle of each element, respectively. 
        Because daily mean soil water is $SW$ and the travel distance of soil water during one day is $LP \cdot \sin(S_{i})$, the daily mean volume of soil water flowing from an element ($V_{SW}$) is $SW \cdot LP \cdot \sin(S_{i}) \cdot w_{i}$.
        Because the interflow produced from the element $i$ ($IF_{i}$) is equal to $V_{SW}$ per unit surface area of the element, $IF_{i}$ should be $\frac{SW \cdot LP \cdot \sin(S_{i}) \cdot w_{i}}{A_{i}}$, and the inflows of subsurface soil water from the contributing area to the element $j$ ($IF(\mathrm{CE})_{j}$) should be $\frac{SW \cdot LP \cdot \sin(S_{i}) \cdot w_{i}}{A_{j}}$.
        The $IF_{i}$ and the $IF(\mathrm{CE})_{j}$ have different values when the surface areas of the elements $i$ and $j$ are different.
    }
    \label{fig:IF_MMMF}
\end{figure}

The MMMF model considers interflow ($IF$; \si{\mm}) as the daily mean amount of subsurface water that flows from an element to downslope elements in one year.
The interflow from upslope elements ($IF(\mathrm{CE})$; \si{\mm}) affects runoff generation processes at an element by reducing the soil moisture storage capacity of the soil ($R_{c}$; \si{\mm}).
The MMMF model uses the following equation to calculate subsurface interflow ($IF_{\mathrm{MMMF}}$; \si{\mm}) from an element (\citet{Morgan2008}, eq. (13)):
\begin{equation}\label{eq:IF_MMMF}
    IF_{\mathrm{MMMF}} = \left(\frac{R-E-Q}{365}\right) \cdot \left( LP \cdot \sin(S) \right), %\tagaddtext{[$\mathrm{mm}\cdot\mathrm{m}/\mathrm{day}$]},
\end{equation}
where $R$ (\si{\mm}) is the mean annual rainfall per unit area, $E$ (\si{\mm}) is the annual evaporation per unit area, $Q$ (\si{\mm}) is the annual runoff per unit area, $LP$ (\si{\m\per\day}) is the saturated lateral permeability as a unit of velocity, $S$ (\si{\degree}) is the slope angle of an element, and 365 is the number of days in one year. 
The first part of equation ~(\ref{eq:IF_MMMF}) corresponds to the daily mean soil water of one year ($SW$; \si{\mm}).
The second part of the equation is the velocity (\si{\m\per\day}) of the interflow of an element, which can be interpreted as the travel distance of interflow during one day (\si{\m}) for daily time steps.
\newline
In the MMMF model, the unit of $IF_{\mathrm{MMMF}}$ is defined as volume per unit area (\si{\liter\per\square\m} = \si{\mm}), which is similar to other hydrological quantities in the model (i.e., $R$, $R\!f$, $R_{c}$, $E$, and $Q$).
However, the unit of interflow in equation~(\ref{eq:IF_MMMF}), which is depth multiplied by velocity (or length for daily time steps), contradicts the definition of $IF_{\mathrm{MMMF}}$ as depth (\si{\mm}) in the model.
\newline
We argue that the interflow equation is improperly formulated and that the $IF_{\mathrm{MMMF}}$ has the wrong unit.
Let us consider the interflow generated from an element $i$, as shown in Figure~\ref{fig:IF_MMMF}.
The daily mean soil water over one year ($SW$) is
\begin{equation}\label{eq:SW}
    SW = \frac{R-E-Q}{365}.
\end{equation}
Because the travel distance of interflow during one day is equal to $LP \cdot \sin(S)$, the volume of interflow during one day is
\begin{equation}
    V_{IF} = SW \cdot LP \cdot \sin(S_{i}) \cdot w_{i}.
\end{equation}
$V_{IF}$ is the depth of soil water ($SW$) multiplied by the travel distance of the interflow ($LP \cdot \sin(S)$) and the width of the element ($w_{i}$), as represented in Figure~\ref{fig:IF_MMMF}.
Equivalent to other hydrological quantities, the quantity of interflow ($IF_{i}$) is the total volume per surface area of an element.
Therefore, $IF_{i}$ can be calculated as
\begin{equation}\label{eq:IF_i}
    IF_{i} = \frac{V_{IF}}{A_{i}} = \frac{SW \cdot LP \cdot \sin(S_{i}) \cdot w_{i}}{A_{i}} = \frac{SW \cdot LP \cdot \sin(S_{i})}{l_{i}}.
\end{equation}
According to equations ~(\ref{eq:SW}) and ~(\ref{eq:IF_i}), interflow ($IF_{\mathrm{corrected}}$) should be calculated as
\begin{equation}
    IF_{\mathrm{corrected}}= \left( \frac{R-E-Q}{365}\right)\cdot \left(LP \cdot \sin(S_{i})\right)\cdot \frac{1}{l_{i}} = \frac{IF_{\mathrm{MMMF}}}{l_{i}}.
\end{equation}
Therefore, the additional term of $\frac{1}{l_{i}}$ is required for the interflow equation of \citet{Morgan2008}.
Furthermore, with this term, $IF_{\mathrm{corrected}}$ has the correct unit of depth (\si{\mm}).
The dependence of interflow on slope length is obvious, as shown in the lateral flow equation of the widely used SWAT model (equation 2:3.5.9 of \citet{SWAT2009}), because the SWAT model also uses water volume per unit area.
We derived the same formula of $IF_{\mathrm{corrected}}$ by using the theoretically well-established Darcy's law in the supplementary material of this article.
\subsubsection{Discrepancy between generated and transferred interflow}
Another problem exists in the interflow equation even if $IF_{\mathrm{corrected}}$ is used rather than $IF_{\mathrm{MMMF}}$.
As shown in Figure~\ref{fig:IF_MMMF}, the generated interflow from the element $i$ flows into the element $j$.
Because the total volume of interflow ($V_{IF}$) is the same for both elements, the interflow into the element $j$ ($IF(\mathrm{CE})_{j}$) should be
\begin{equation}
    IF(\mathrm{CE})_{j} = \frac{V_{IF}}{A_{j}} \neq \frac{V_{IF}}{A_{i}} = IF_{i}.
\end{equation}
The discrepancy between generated and transferred interflow is attributed to the different surface areas of the elements. 
If using raster maps in the MMMF models (Figure~\ref{fig:IF_MMMF}), as is performed in most MMF model studies, the extent of the discrepancy can be calculated as
\begin{equation}
    \frac{IF_{i}}{IF(\mathrm{CE})_j} = \frac{V_{IF}}{A_{i}} \cdot \frac{A_{j}}{V_{IF}} = \frac{l_{j}}{l_{i}} = \frac{\cos(S_{i})}{\cos(S_{j})}.
\end{equation}
Therefore, the discrepancy is larger if the difference in slope between adjacent upslope and downslope elements is significant. 
Similar discrepancies between adjacent elements can also be found in every matter exchange processes of the MMMF model (i.e., surface runoff, interflow, and sediment). 
Problems of the discrepancy can be solved by using the water volume or the total sediment mass for transferring water and sediments and dividing the volume or total mass by the surface area of the receiving element.
\subsubsection{Consequence of the error in calculating interflow}
\begin{figure}[H]
    \centering
    \includegraphics{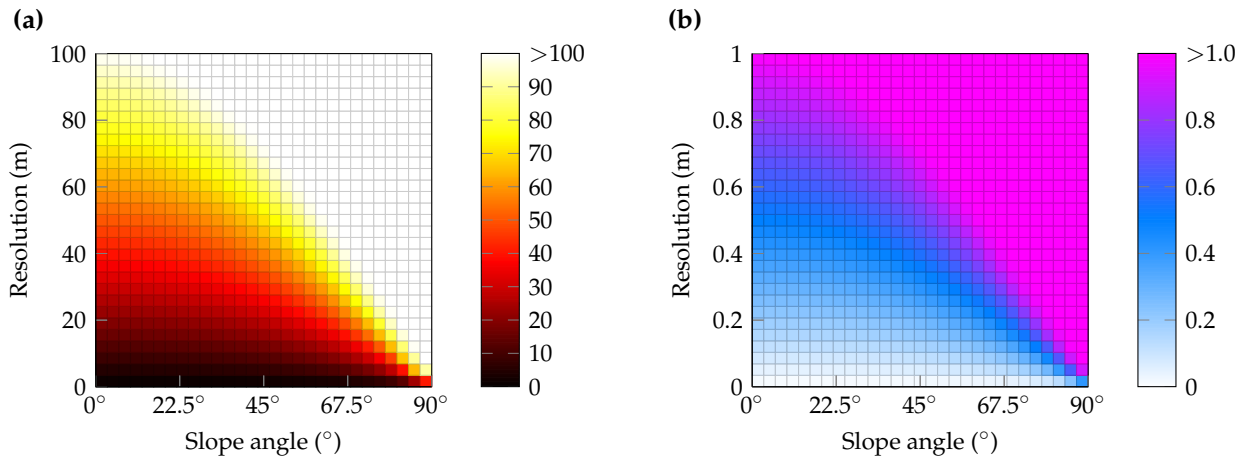}
    \caption{Extent of the overestimation of $IF_{\mathrm{MMMF}}$ compared with $IF_{\mathrm{corrected}}$ as a fraction of $\frac{IF_{\mathrm{MMMF}}}{IF_{\mathrm{corrected}}}$ by slope and resolution.
    \textbf{(a)} General pattern for maximum resolution of \SI{100}{\m}.
    \textbf{(b)} Pattern for fine-resolution section.
    Values larger (smaller) than one indicate overestimation (underestimation).
    }
    \label{fig:IFComp}
\end{figure}

\begin{figure}[H]
    \centering
    \includegraphics{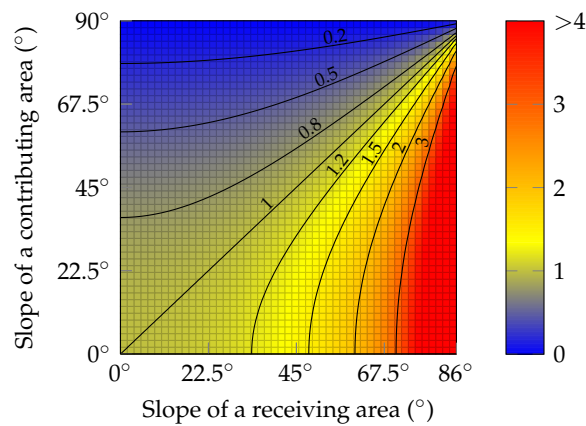}
    \caption{Extent of the discrepancy between the generated interflow from a contributing element and the transferred interflow to a receiving element. 
    }
    \label{fig:IFComp_ij}
\end{figure}

Owing to the incorrect formula in the interflow equation of \citet{Morgan2008}, the MMMF model overestimates interflow when the slope length of an element increases. 
However, the model underestimates interflow if the slope length of an element is less than \SI{1}{\m}.
Figure~\ref{fig:IFComp} shows the extent of the overestimation when the MMMF model is applied to raster maps such as a digital elevation model (DEM).
For a DEM with a certain resolution ($res$; \si{\m}), the width ($w$) and the length ($l$) of each element are equal to $res$ and $\frac{res}{\cos(S)}$.
Therefore, the extent of overestimation of the interflow is dependent on the slope of an element and the resolution of the DEM.
In the case of \citet{Setiawan2012}, who applied the MMMF model using a DEM with \SI{0.05}{\m} resolution for a maximum slope of 41$^\circ$, the MMMF model estimated at most only \SI{7}{\percent} of the corrected interflow.  
In the case of \citet{Lilhare2014}, who applied the MMMF model using a DEM with \SI{90}{\m} resolution, the extent of overestimation was more than 90 times compared with the corrected interflow.
\newline
Owing to the discrepancy between the generated and transferred interflow, the model overestimates the interflow from contributing elements to a receiving element when the receiving element is steeper than the contributing elements, as shown in Figure~\ref{fig:IFComp_ij}.
In addition, the model underestimates the interflow from contributing elements when they are steeper than the receiving element.
The extent of the discrepancy increases with the increase in slope differences between contributing and receiving elements.
Because errors in the calculation of interflow are positively correlated with the size, slope, and rate of change in the slope of elements, the model is not suitable for steep mountainous terrain with complex topography.
Furthermore, the interflow affects the quantity of surface runoff by affecting the amount of the soil moisture storage ($R_{c}$), as shown in eq. (10) of \citet{Morgan2008}. 
\subsection{Improperly normalized C-factor in the transport capacity equation}
The MMF model uses a crop cover management factor (C-factor), which is the ratio of soil loss under a given surface condition (actual condition) to that from a bare ground condition (reference condition) based on the empirical values of Universal Soil Loss Equation (USLE) cropping (C) and erosion control (P) factors \citep{Morgan1984}.
In the MMMF model, \citet{Morgan2008} the C-factor is calculated by modifying the ratio of soil loss to the ratio of runoff velocity under an actual condition as that of the reference condition.
This modification allows the MMMF model to generalize the empirically based C-factor by using measurable physical quantities. 
As a result, the model can consider the effects of surface roughness, rill depth, and vegetation structure on soil erosion in addition to the effect of crop cover management on soil erosion.
However, we argue that the C-factor used in the MMMF model is not properly normalized in the course of combining multiple velocities (i.e., $v_a$, $v_v$, and $v_t$) corresponding to different surface condition types.
As a consequence, if the model considers more than one surface condition type, the unitless C-factor contains the inconsistent units of the velocity or the squared value of velocity.
The MMMF model calculates the C-factor as (eqs. (39)--(41) of \citet{Morgan2008})
\begin{equation}
    \mathrm{C_{MMMF}} = \frac{v_a \cdot v_v \cdot v_t}{v_b}, 
\end{equation}
where $v_a$, $v_v$, and $v_t$ are runoff velocity considering the rill condition, vegetation cover, and surface roughness, respectively.
$v_b$ is the runoff velocity for the reference condition of unchanneled overland flow over smooth bare ground.
Moreover, adding and subtracting variables are allowed in the MMMF model, according to the surface condition.
Assuming that only one of the surface condition types is considered in the model, the C-factor for each surface condition type should be calculated as 
\begin{align}
    \mathrm{C}_{a} &= \frac{v_a}{v_b}\\
    \mathrm{C}_{v} &= \frac{v_v}{v_b}\\
    \mathrm{C}_{t} &= \frac{v_t}{v_b}.
\end{align}
As described in the MMMF model, all surface condition types are considered independently from each other, which means that a surface condition type is not affected by other surface condition types.
Therefore, for the combination of the surface condition types, the C-factor should be calculated as
\begin{equation}\label{eq:C_corr}
    \mathrm{C_{corrected}} = \frac{v_a}{v_b} \cdot \frac{v_v}{v_b} \cdot \frac{v_t}{v_b} = \frac{v_a \cdot v_v \cdot v_t}{{v_b}^{3}}.
\end{equation}
According to equation ~(\ref{eq:C_corr}), $\mathrm{C_{corrected}}$ is unitless because each velocity is normalized by the reference velocity. 
Even if some surface condition types are missing or added, the unit of the factor remains constant.
\subsubsection{Consequence of improper normalization of the C-factor}
According to \citet{Petryk1975}, rill depth (hydraulic radius) acts as an accelerator of the runoff velocity, whereas vegetation and surface roughness act as resistors of the runoff velocity.
Therefore, the C-factor should be increased when the model additionally considers a surface with a rill depth deeper than that of reference surface condition (\SI{0.005}{m}).
On the contrary, the C-factor should be decreased when the model additionally considers vegetation cover and surface roughness of an element.
However, for hillslopes in which $v_a$, $v_v$, and $v_t$ are faster than \SI{1}{\m\per\s}, the C-factor of the MMMF model sharply increases by a factor of the added runoff velocity even if the model additionally considers vegetation cover or surface roughness.
For slopes with runoff velocities lower than \SI{1}{\m\per\s}, the C-factor is underestimated when the model additionally considers surface condition type. 
Errors occur because the C-factor of the MMMF model does not consider the relative velocity of the reference surface condition. 
The effect of the error is significant for elements with high runoff velocities when the soil erosion rates are high.
Owing to the slope dependence of runoff velocity, increased slope of an element relates to greater overestimation.
\section{Conclusions}
The MMF model is a widely used semi-physically based soil erosion model because it includes rigorous physical processes, easily understood features, and moderate data requirements.
The newly added features of \citet{Morgan2008} consider the slope angle, subsurface water processes, surface conditions, and characteristics of each soil particle type \citep{Lilhare2014}, which have the potential to further strengthen the physical basis of the model.
We identified three problematic formulations related to the calculations of effective rainfall, interflow, and the C-factor of transport capacity, which can produce inadequate results of runoff and soil erosion.
In addition, we suggested alternative formulations to provide more accurate estimations of runoff and soil erosion.
\vspace{6pt} 

%%%%%%%%%%%%%%%%%%%%%%%%%%%%%%%%%%%%%%%%%%
\acknowledgments{This study is part of the International Research Training Group ``Complex Terrain and Ecological Heterogeneity'' (TERRECO) (GRK 1565/1) funded by the German Research Foundation (DFG). The document was typed by \LaTeX{} using the modified class and bibliography style files provided by MDPI (\url{http://www.mdpi.com/authors/latex}). The authors would like to thank Jean-Lionel Payeur-Poirier and Anthony from Editage for English proofreading this manuscript.}
%%%%%%%%%%%%%%%%%%%%%%%%%%%%%%%%%%%%%%%%%%
\bibliographystyle{arxiv}

\bibliography{Bayreuth.bib}

\pagebreak
\begin{center}
\textbf{\large Supplemental Material}
\end{center}
\setcounter{section}{0}
\setcounter{equation}{0}
\setcounter{figure}{0}
\setcounter{table}{0}
\makeatletter
\renewcommand{\thesection}{S\arabic{section}}
\renewcommand{\theequation}{S\arabic{equation}}
\renewcommand{\thefigure}{S\arabic{figure}}
\renewcommand{\bibnumfmt}[1]{[S#1]}
\renewcommand{\citenumfont}[1]{S#1}
\section{Derivation of the corrected interflow equation using Darcy's law}
\begin{figure}[H]
\centering
    \includegraphics{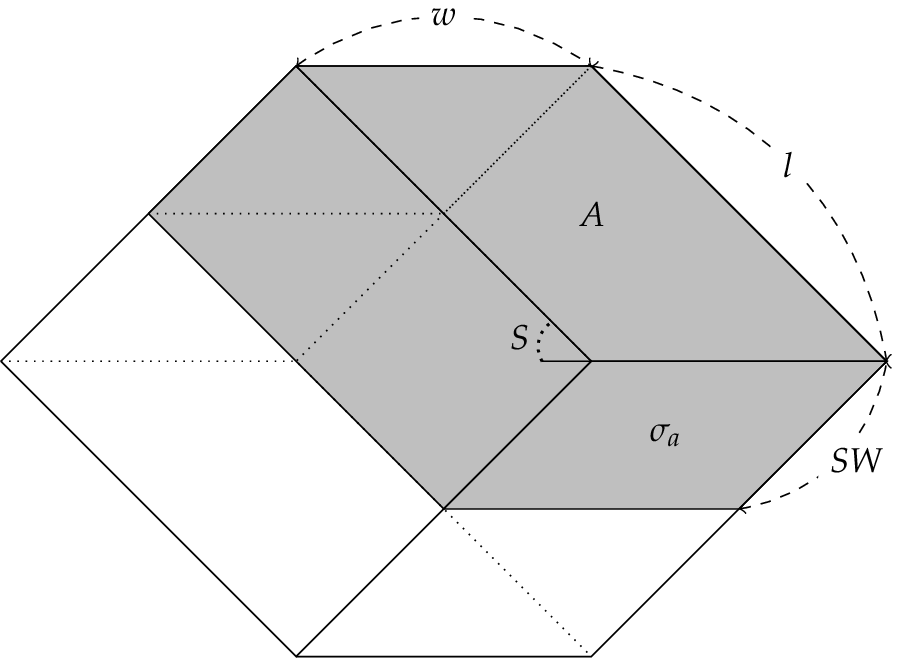}
    \caption{Conceptual representation of soil water in an element.
    $A$, $w$, $l$, and $S$ are the area (\si{\square\m}), width (\si{\m}), slope length (\si{\m}), and slope angle (\si{\degree}) of the element, respectively; $SW$ (\si{\mm}) is the daily mean soil water over one year, and $\sigma_{a}$ (\si{\square\m}) is the cross-sectional area of $SW$. 
    }
    \label{fig:S1}
\end{figure}

Let us assume that there is an amount of soil water equivalent to the daily mean soil water over one year ($SW$; \si{\mm}) in an element and that only soil water exerts a force on the element for the interflow process (Figure~\ref{fig:S1}).
The volumetric flux of soil water ($J_{SW}$; \si{\cubic\m\per\s}) in the element can be derived from Darcy's law when the hydraulic conductivity ($K$) of the soil is given: 
\begin{equation}\label{eq:J}
    J_{SW} = -\frac{K \cdot \sigma_{a}}{\rho \cdot g} \cdot \frac{\Delta P}{l},
\end{equation}
where $\sigma_{a}$ (\si{\square\m}) is the cross-sectional area of the soil water along the downslope direction, $\rho$ (\si{\kg\per\cubic\m}) is the density of the soil water, $g$ (\si{\m\per\square\s}) is the gravitational acceleration, $\Delta P$ (\si{\pascal}) is a pressure gradient of the soil water for both sides of the element, and $l$ (\si{\m}) is the length of the soil water along the downslope direction. 
Because pressure ($P$) is force ($f$; \si{\newton}) divided by the cross-sectional area, $P$ can be calculated as
\begin{equation}\label{eq:P}
    P = \frac{f}{\sigma_{a}}.
\end{equation}
The gradient of force ($\Delta f$; N) acting on both ends of the element can be derived from the volume of the soil water ($V'_{SW}$; \si{\cubic\m}), which is calculated as
\begin{equation}\label{eq:V}
    V'_{SW} = 0.001 \cdot SW \cdot A = 0.001 \cdot SW \cdot w \cdot l, 
\end{equation}
where $A$ (\si{\square\m}), $w$ (\si{\m}), and $l$ (\si{\m}) are the surface area, width, and length of the element, respectively.
The unit conversion factor of 0.001 is used to convert millimeters to meters.
Because the gradient of force depends on the gradient of the mass of both ends of the element, $\Delta f$ is calculated as
\begin{equation}\label{eq:df}
    \Delta f = V'_{SW} \cdot \rho \cdot g \cdot \sin{(S)},
\end{equation}
where $S$ is the slope angle (\si{\degree}) of the element.
From equations ~(\ref{eq:J}), ~(\ref{eq:P}), ~(\ref{eq:V}), and ~(\ref{eq:df}), the volumetric flux of the soil water can be simplified as
\begin{equation}\label{eq:Jsimple}
    J_{SW} = -0.001 \cdot K \cdot SW \cdot w \cdot \sin{(S)}.
\end{equation}
Because the hydraulic conductivity during one day is defined as saturated lateral permeability ($LP$; $\mathrm{m/day}$) in the MMMF model, the volume of soil water flowing from the element ($V_{SW.out}$) during one day is
\begin{equation}\label{eq:V_out}
    V_{SW.out} = J_{SW} \cdot \mathrm{1\enspace (day)} = \left(0.001 \cdot LP \cdot SW \cdot w \cdot \sin{(S)}\right) \cdot \mathrm{1\enspace (day)}.
\end{equation}
Because the time step of one day affects only the unit of equation ~(\ref{eq:V_out}), the interflow from the element ($IF_{\mathrm{corrected}}$; \si{\mm}) as a quantity of volume per unit surface area can be calculated as
\begin{equation}
    IF_{\mathrm{corrected}} = 1000 \cdot \frac{V_{SW.out}}{A} = \frac{LP \cdot SW \cdot w \cdot \sin{(S)}}{l},
\end{equation}
where 1000 is the unit converting factor from meters to millimeters.
Therefore, we can use Darcy's law to obtain an identical equation as that in the main text that depends on slope length.
\end{document}